\documentclass[dvips]{article}
\usepackage{amsmath}
\usepackage{amssymb}
\usepackage{graphicx}
\usepackage{epsfig}
\usepackage{lineno}
\usepackage{multirow}

\title{The effect of the geomagnetic field on cosmic ray energy estimates and large scale anisotropy searches on data from the Pierre Auger Observatory}
\date{}

\begin{document}
\maketitle

\par\noindent
{\bf The Pierre Auger Collaboration} \\
P.~Abreu$^{74}$, 
M.~Aglietta$^{57}$, 
E.J.~Ahn$^{93}$, 
I.F.M.~Albuquerque$^{19}$, 
D.~Allard$^{33}$, 
I.~Allekotte$^{1}$, 
J.~Allen$^{96}$, 
P.~Allison$^{98}$, 
J.~Alvarez Castillo$^{67}$, 
J.~Alvarez-Mu\~{n}iz$^{84}$
% ...

\par\noindent
$^{1}$ Centro At\'{o}mico Bariloche and Instituto Balseiro (CNEA-
UNCuyo-CONICET), San Carlos de Bariloche, Argentina \\
% ...
$^{19}$ Universidade de S\~{a}o Paulo, Instituto de F\'{\i}sica, S\~{a}o 
Paulo, SP, Brazil \\
% ...
$^{33}$ Laboratoire AstroParticule et Cosmologie (APC), 
Universit\'{e} Paris 7, CNRS-IN2P3, Paris, France \\
% ...
$^{57}$ Istituto di Fisica dello Spazio Interplanetario (INAF),
 Universit\`{a} di Torino and Sezione INFN, Torino, Italy \\
% ...
$^{67}$ Universidad Nacional Autonoma de Mexico, Mexico, D.F., 
Mexico \\
% ...
$^{74}$ LIP and Instituto Superior T\'{e}cnico, Technical 
University of Lisbon, Portugal \\
% ...
$^{84}$ Universidad de Santiago de Compostela, Spain \\
% ...
$^{93}$ Fermilab, Batavia, IL, USA \\
% ...
$^{96}$ New York University, New York, NY, USA \\
% ...
$^{98}$ Ohio State University, Columbus, OH, USA \\
% ...
\par\noindent
($\dagger$) Deceased \\
(a) at Konan University, Kobe, Japan \\
(f) now at University of Maryland \\

\begin{abstract}
We present a comprehensive study of the influence of the geomagnetic 
field on the energy estimation of extensive air showers with a zenith
angle smaller than $60^\circ$, detected at the Pierre Auger Observatory.
The geomagnetic field induces an azimuthal modulation of the estimated energy of cosmic rays up to the $\sim2\%$ level at large zenith angles.
We present a method to account for this modulation of the reconstructed 
energy. We analyse the effect of the modulation on
 large scale anisotropy searches in the arrival direction distributions of cosmic rays.  
At a given energy, the geomagnetic effect is shown to induce a pseudo-dipolar pattern
 at the percent level in the declination distribution that needs to be accounted
 for.
\end{abstract}

\vfill\eject

\section{Introduction}

High energy cosmic rays generate extensive air showers in the 
atmosphere. The trajectories of the charged particles of the showers 
are curved in the Earth's magnetic field, resulting in a broadening 
of the spatial distribution of particles in the direction of the 
Lorentz force. While such effects are known to distort the particle 
densities in a dramatic way at zenith angles larger than 
$\sim$60$^\circ$~\cite{edge,ave,ave2,dembinski}, they are commonly 
ignored at smaller zenith angles where the lateral distribution 
function is well described by empirical models of the 
NKG-type~\cite{greisen,kamata} based on a radial symmetry of the 
distribution of particles in the plane perpendicular to the shower 
axis. 

In this article, we aim to quantify the small changes of the
particle densities at ground induced by the geomagnetic field
for showers with zenith angle smaller than $\sim$60$^\circ$, 
focusing on the impacts on the energy estimator used at the
Pierre Auger Observatory. As long as the magnitude of these
effects lies well below the statistical uncertainty of the energy 
reconstruction, it is reasonable to neglect them in the framework 
of the energy spectrum reconstruction. As the strength 
of the geomagnetic field component perpendicular to the arrival
direction of the cosmic ray, $B_{\rm T}$, 
depends on both the zenith and the azimuthal angles 
$(\theta,\varphi$) of any incoming shower, these effects are 
expected to break the symmetry of the energy estimator in terms 
of the azimuthal angle $\varphi$. Such an azimuthal dependence 
translates into azimuthal 
modulations of the estimated cosmic ray event rate at a given energy. 
For any observatory located far from the Earth's poles, any genuine large scale pattern which depends on the declination
translates also into azimuthal modulations of the cosmic ray event 
rate. Thus to perform a large scale anisotropy measurement it is 
critical to account for azimuthal modulations of
experimental origin and for those induced by the geomagnetic field,
as already pointed out in the analysis of the Yakutsk data~\cite{ivanov} and the ARGO-YBJ data~\cite{argoybj}.
Hence, this work constitutes an accompanying paper of a search for
large scale anisotropies, both in right ascension and declination
of cosmic rays detected at the Pierre Auger Observatory, the results of which will be
reported in a forthcoming publication. 

To study the influence of the geomagnetic field on the cosmic ray 
energy estimator, we make use of shower simulations and of the measurements performed with the surface 
detector array of the Pierre Auger Observatory, located in Malarg\"{u}e, 
Argentina (35.2$^\circ$S, 69.5$^\circ$W) at 1400~m a.s.l.~\cite{auger1}. 
The Pierre Auger Observatory is designed to study cosmic rays (CRs)
with energies above 
$\sim 10^{18}\,$eV. The surface detector 
array consists of 1660 water Cherenkov detectors sensitive to the 
photons and the charged particles of the showers. It is laid out 
over an area of 3000~km$^2$ on a triangular grid and is 
overlooked by four fluorescence detectors. The energy at which
the detection efficiency of the surface detector array saturates 
is $\sim 3\,$EeV~\cite{auger2}. For each event, the signals recorded 
in the stations are fitted to find the signal at $1000$~m from the 
shower core, $S(1000)$, used as a measure of the shower size. The 
shower size $S(1000)$ is converted to the value $S_{38}$ that would have 
been expected had the shower arrived at a zenith angle of 38$^\circ$.
$S_{38}$ is then converted into energy using a calibration curve based on the fluorescence 
telescope measurements~\cite{auger3}. 

The influence of the geomagnetic field on the spatial distribution 
of particles for showers with zenith angle less than 60$^\circ$ is
presented in Section~\ref{sec:geom}, through a toy model aimed at 
explaining the directional dependence of the shower size $S(1000)$ 
induced by the geomagnetic field. The observation of 
this effect in the data of the Pierre Auger Observatory is reported in Section~\ref{sec:obs}. In 
Section~\ref{sec:azim}, we quantify the size of the $S(1000)$ 
distortions with zenith and azimuthal angles by means of end-to-end shower 
simulations, and then present the procedure to convert the shower size 
corrected for the geomagnetic effects into energy using the 
Constant Intensity Cut method. In Section~\ref{sec:aniso}, the 
consequences on large scale anisotropies are discussed, while
systematic uncertainties associated with the primary mass, the primary 
energy and the number of muons in showers are presented in 
Section~\ref{sec:syst_mod}.

\section{Influence of the geomagnetic field on extensive air showers}
\label{sec:geom}

The interaction of a primary cosmic ray in the atmosphere produces mostly charged and neutral pions, initiating a hadronic cascade.
The decay of neutral pions generates the
electromagnetic component of the shower, while the decay of the
charged pions generates the muonic one. Electrons undergo stronger scattering, so that the electron
distribution is only weakly affected by the geomagnetic deflections. Muons are produced with
a typical energy $E_\mu$ of a few GeV (increasing with the altitude 
of production). The decay angle between pions and muons
is causing only a small additional random deflection, as they almost 
inherit the transverse momentum $p_{\rm T}$ of their parents (a few 
hundred MeV/$c$) so that the distance of the muons from the shower core
scales as the inverse of their energy. While the radial 
offset of the pions from the shower axis is of the order of a few 
10~m, it does not contribute significantly to the lateral distribution 
of the muons observed on the ground at distances $r\geq100\,$m. 
Hence, at ground level, the angular spread of the muons around 
the shower axis can be considered as mainly caused by the transverse 
momentum inherited from the parental pions.

After their production, muons are affected by ionisation and 
radiative energy losses, decay, multiple scattering and geomagnetic
deflections. Below 100~GeV, the muon energy loss is mainly due to
ionisation and is relatively small (amounting to about 
2~MeV~g$^{-1}$~cm$^2$), allowing a large fraction of muons to
reach the ground before decaying. Multiple scattering in the 
electric field of air nuclei randomises the directions of muons
to some degree, but the contribution to the total angular divergence
of the muons from the shower axis remains small
up to zenith angles of the shower-axis of about $80^\circ$. 

\begin{figure}[!t]
  \centering					 
  \includegraphics[width=8cm]{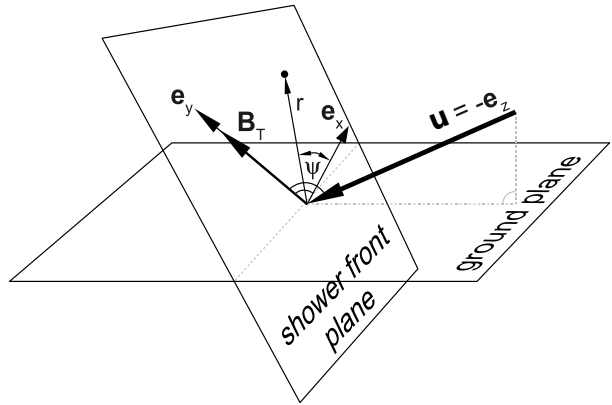}
  \caption{\small{The shower front plane coordinate system~\cite{ave,dembinski}:
\textbf{e}$_z$ is anti-parallel to the shower direction
$\mathbf{u}$, while \textbf{e}$_y$ is parallel to \textbf{B}$_{\rm T}$, the projection of the magnetic field $\mathbf{B}$ onto the shower plane x-y. $(\psi,r)$ are the polar coordinates in the shower plane.}}
  \label{coordinatesystem}
\end{figure}

Based on these general considerations, we now introduce a simple 
toy model aimed at understanding the main features of the muon 
density distortions induced by the geomagnetic field. We adopt
the shower front plane coordinate system depicted in 
Fig.~\ref{coordinatesystem}~\cite{ave}.
In the absence of the magnetic field, and neglecting multiple scattering,
a relativistic muon of energy $E_\mu\simeq cp_\mu$ and transverse
momentum $p_{\rm T}$ will reach the shower front plane after traveling a 
distance $d$ at a position $r$ from the shower axis given by
\begin{equation}
\label{eqn:r}
r\simeq\frac{p_{\rm T}}{p_\mu}\,d\simeq\frac{cp_{\rm T}}{E_\mu}\,d.
\end{equation}
On the other hand, in the presence of the magnetic field, muons suffer
additional geomagnetic deflections. We treat the geomagnetic 
field \textbf{B} in Malarg\"{u}e as a constant field\footnote{In Malarg\"{u}e the 
geomagnetic field has varied by about 1$^\circ$ in direction 
and 2\% in magnitude over 10 years~\cite{natgeo}.},
\begin{equation}
\label{eqn:b}
B=24.6\,\mu\mathrm{T},\hspace{1cm}D_\mathrm{B}=2.6^\circ,\hspace{1cm}I_\mathrm{B}=-35.2^\circ,
\end{equation}
$D_\mathrm{B}$ and $I_\mathrm{B}$ being the geomagnetic declination and inclination. The deflection of a relativistic muon in the presence of a magnetic field with transverse component $B_{\rm T}$ can be approximated with 
\begin{equation}
\label{eqn:delta}
\delta x_\pm\simeq\pm\frac{ecB_{\rm T}d^2}{2E_\mu},
\end{equation}
\begin{figure}[!t]
  \centering					 
  \includegraphics[width=8cm]{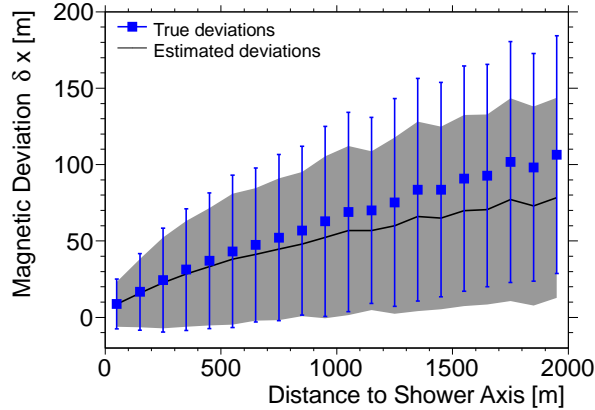}
  \caption{\small{Magnetic deviations as a function of the distance to
  the shower axis observed on a simulated vertical shower (points). Superimposed 
  are the deviations expected from Eq.~\eqref{eqn:delta} (line). The shaded region and the error bars give the corresponding dispersion.}} 
  \label{averagedev}
\end{figure}
where $e$ is the elementary electric charge and the sign corresponds to positive/negative charged muons. 
The dependence of the geomagnetic deflections $\delta x \equiv \delta x_{+} = - \delta x_{-}$ on the distance
to the shower axis $r=\sqrt{x^2+y^2}$ is illustrated
in Fig.~\ref{averagedev} obtained by comparing the position of the
same muons in the presence or in the absence of the geomagnetic 
field in a simulated vertical shower of a proton at 5~EeV. 
The deviations expected from the expression for $\delta x_\pm$ are also 
shown in the same graph (solid line). It was obtained by inserting muon energy and distance at the production point of the simulated muons into Eq.~\eqref{eqn:delta}. 
It turns out that Eq.~\eqref{eqn:delta} estimates 
rather well the actual deviations, though the distance between the actual 
and the predicted deviations increases at large $r$. This is mainly because on the one hand 
$d$ underestimates the actual travel length to a larger extent at larger 
$r$, while on the other hand the magnetic deviation actually increases
while muons gradually lose energy during travel.
Hence, from the muon density $\rho_\mu(x,y)$ in the transverse plane 
in the absence of the geomagnetic field, the corresponding density 
$\overline{\rho}_\mu(\overline{x},\overline{y})$ in the presence of such a field can be obtained by making the following Jacobian 
transformation, in the same way as in the framework of very inclined
showers~\cite{ave},
\begin{equation}
\label{eqn:rho}
\overline{\rho}_\mu(\overline{x},\overline{y})=\left|\frac{\partial{(x,y)}}{\partial{(\overline{x},\overline{y})}}\right|\,\rho_\mu(x(\overline{x},\overline{y}),y(\overline{x},\overline{y})).
\end{equation}
Here, the term ``muon density'' refers to the time-integrated muon
flux through the transverse shower front plane associated to the air shower, 
and the barred coordinates represent the positions of the muons
in the transverse plane in the presence of the geomagnetic field:
\begin{eqnarray}
\label{eqn:xy}
\overline{x}&=&x+\delta x_\pm(x,y),\nonumber\\
\overline{y}&=&y.
\end{eqnarray}
Since Eq.~\eqref{eqn:rho} induces changes of the shower size $S(1000)$,
it is of particular interest to get an approximate relationship between
$\rho$ and $\overline{\rho}$ around $1000$~m. From Fig.~\ref{averagedev}, it is apparent that around 
$1000$~m the mean magnetic deviation is approximately constant over a 
distance range larger than the size of the deviation. When focusing on 
the changes of density at $1000$~m from the shower core, it is thus 
reasonable to neglect the $x$ and $y$ dependence of the deviation 
$\delta x_\pm$, which allows an approximation of the density 
$\overline{\rho}_\mu(\overline{x},\overline{y})$ around $1000$~m as
\begin{eqnarray}
\label{eqn:rho2}
\overline{\rho}_\mu(\overline{x},\overline{y})&\simeq&\rho_{\mu_+}(\overline{x}-\delta x_+,\overline{y})+\rho_{\mu_-}(\overline{x}-\delta x_-,\overline{y})\nonumber\\
&\simeq&\rho_\mu(\overline{x},\overline{y})+\frac{(\delta x)^2}{2}\frac{\partial^2\rho_\mu}{\partial \overline{x}^2}(\overline{x},\overline{y}),
\end{eqnarray}
where we assumed 
$\rho_{\mu_-}=\rho_{\mu_+}=\rho_{\mu}/2$. The two opposite muon charges 
cancel out the linear term in $\delta x$ and we see that magnetic effects change the muon density around $1000$~m by a 
factor proportional to $(\delta x)^2 \propto B_{\rm T}^2 \propto \sin^2(\widehat{\textbf{u},\textbf{b}})$, where $\textbf{u}$ and $\textbf{b}=\textbf{B}/|B|$ denote 
the unit vectors in the shower direction and the magnetic field direction, respectively. This is particularly important 
with regard to the azimuthal behaviour of the effect, as the
azimuthal dependence is contained \emph{only} in the $B_{\rm T}^2(\theta,\varphi)$ term. 
This dependency is therefore a generic expectation outlined by this toy 
model. The model will be verified in Section~\ref{sec:azim} by making use of 
complete simulation of showers. On the other hand, the zenith angle 
dependence relies on other ingredients that we will probe in an accurate 
way in Section~\ref{sec:azim}, such as the altitude distribution of the muon 
production and the muon energy distribution. 

\section{Observation of geomagnetic effects in the Pierre Auger Observatory data}
\label{sec:obs}

\begin{figure}[!t]
  \centering					 
  \includegraphics[width=8cm]{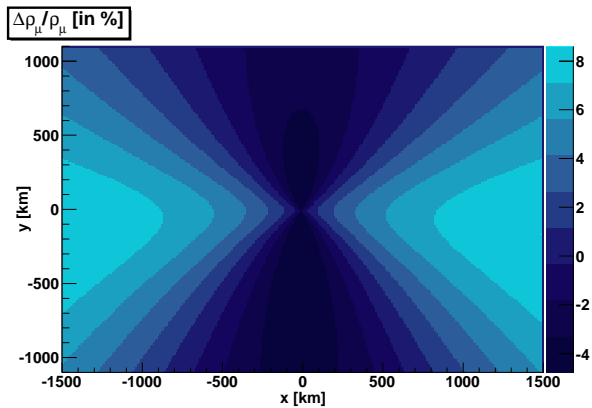}
  \caption{\small{Relative changes of $\Delta\rho_\mu/\rho_\mu$ 
in the transverse shower front plane due to the presence of the 
geomagnetic field, obtained at zenith angle $\theta=60^\circ$ and 
azimuthal angle aligned along $D_{\rm B}+180^\circ$.}}
  \label{mumap}
\end{figure}

\begin{figure}[!t]
  \centering					 
  \includegraphics[width=8cm]{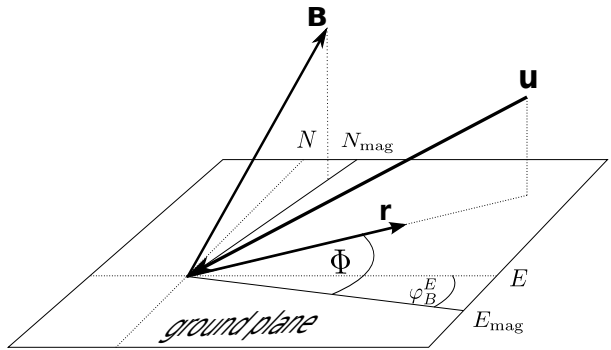}
  \caption{\small{Definition of angle $\Phi$ with respect to the magnetic East $E_{\mathrm{mag}}$ and the shower core for a given shower direction $\mathbf{u}$ and a surface detector at $\mathbf{r}$. The azimuthal angle of the magnetic field vector $\mathbf{B}$ defines the magnetic North  $N_{\mathrm{mag}}$.}}
  \label{groundcs}
\end{figure}

\begin{figure}[!t]
  \centering					 
  \includegraphics[width=7cm]{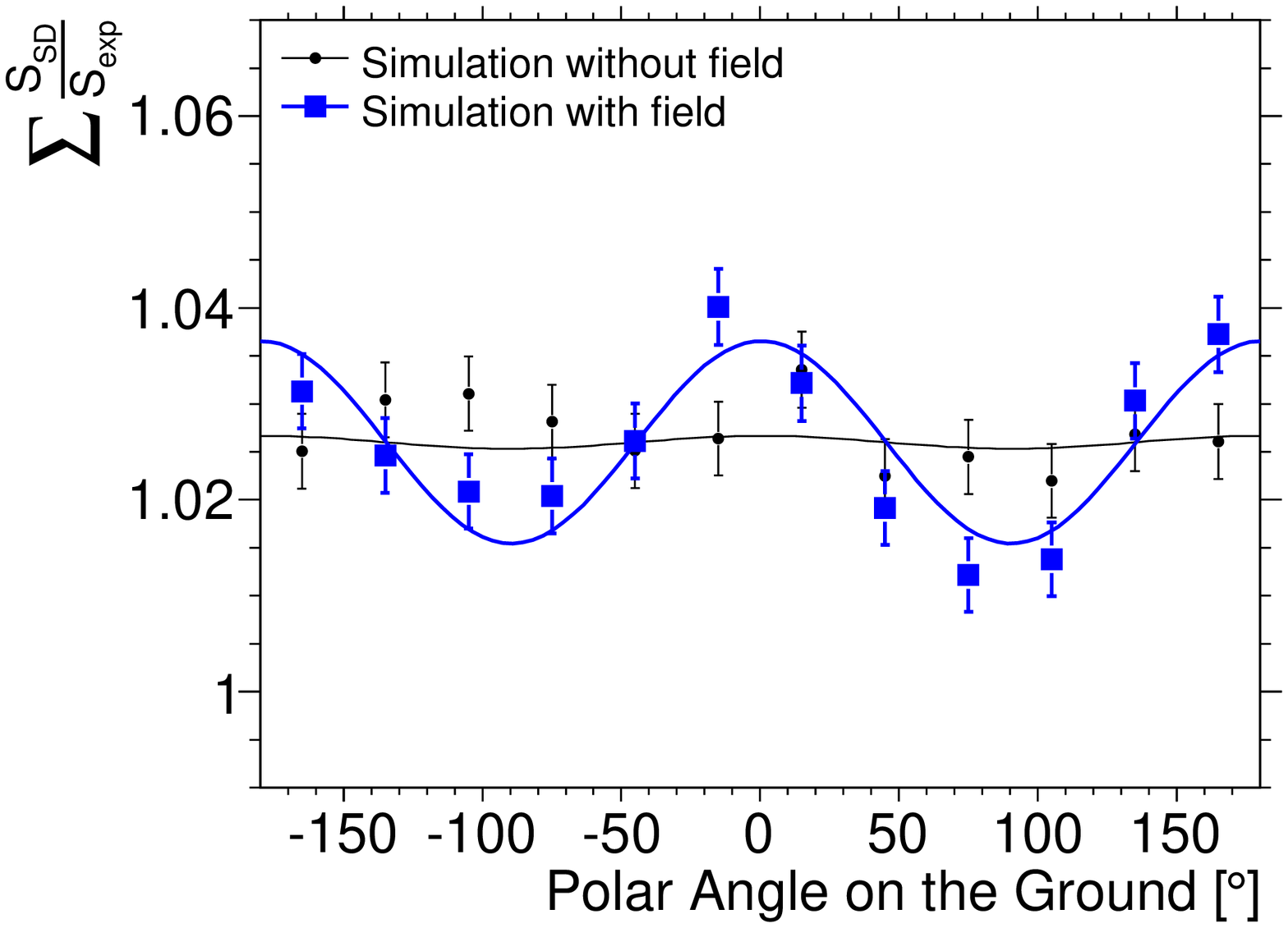}
  \includegraphics[width=7cm]{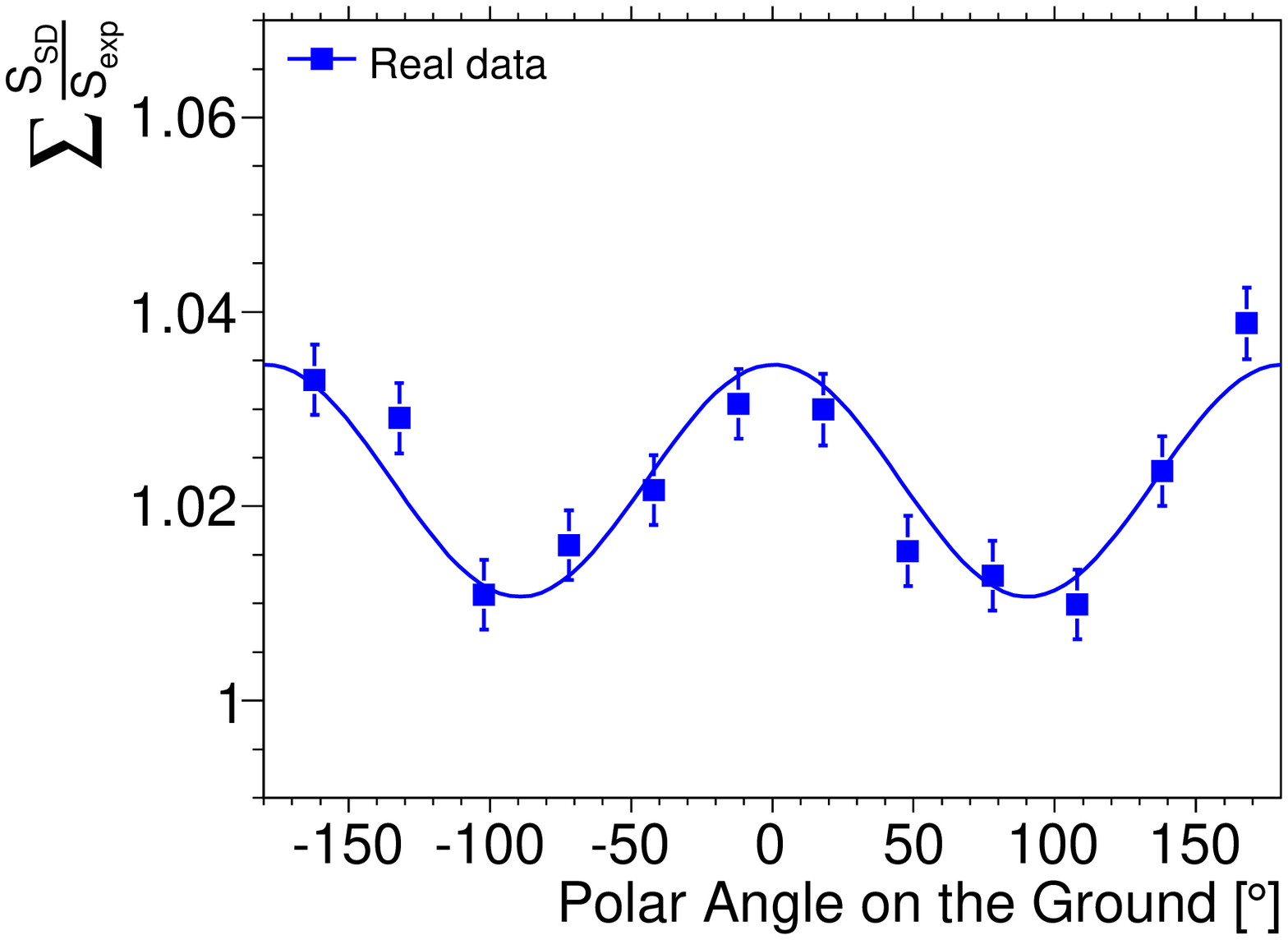}
  \caption{\small{Average ratio of the true signal in each surface detector 
with respect to the expected one as a function of the polar angle on 
the ground. Left panel: using simulated showers in the presence 
(thick points) and in the absence (thin points) of the geomagnetic field. Right 
panel: using real data above 4~EeV. The solid lines give the fit of a quadrupolar modulation to the corresponding points.}}
  \label{tanksig}
\end{figure}

To illustrate the differences between $\overline{\rho}_\mu$ and
$\rho_\mu$ described in Eq.~\eqref{eqn:rho}, the relative changes 
$\Delta\rho_\mu/\rho_\mu$ are shown in Fig.~\ref{mumap} in the 
transverse shower front plane by producing muon maps from simulations 
at zenith angle $\theta=60^\circ$ and azimuthal angle aligned along 
$D_\mathrm{B}+180^\circ$ in the presence and in the absence of the geomagnetic field. 
A predominant quadrupolar asymmetry at the few percent level is 
visible, corresponding to the separation of positive and negative 
charges in the direction of the Lorentz force. 

This quadrupolar asymmetry is expected to induce to some extent a 
quadrupolar modulation of the surface detector signals as a function of 
the \emph{polar angle on the ground}, defined here as the angle between the axis given by the shower core and the surface detector, and the magnetic East $\varphi_\mathrm{B}^\mathrm{E}=-D_\mathrm{B}=-2.6^\circ$ (Fig.~\ref{groundcs}). 
The use of this particular angle, instead of the polar angle 
$\psi$ which is defined in the \emph{shower front plane} (see Fig.~\ref{coordinatesystem}), allows us to remove 
dipolar asymmetries in the surface detector signals, the origin of which is related to the
radial divergence of particles from the shower axis. Such
asymmetries cancel out in this analysis, due to the isotropic distribution of the cosmic rays.
To demonstrate the geomagnetic effect, we produced a realistic 
Monte-Carlo simulation using $30\,000$ 
isotropically distributed showers (with zenith angles less than 60$^\circ$) 
with random core positions within the 
array. The injected primary energies were chosen to be greater than 4~EeV (safely excluding angle dependent trigger probability) and distributed according to a 
power law energy spectrum $dN/dE \propto E^{-\gamma}$ with power index $\gamma=2.7$, so that this 
shower library is as close as possible to the real data set. To each shower we apply the reconstruction procedure of the surface detector, leading to a fit of the lateral distribution function~\cite{auger3}. The lateral distribution function parametrizes the signal strength in the shower plane, assuming circular shower symmetry. By evaluating the lateral distribution function at the position of the surface detector, we obtain the expected signal $S_{\mathrm{exp}}$. This signal can be compared to the true signal in the surface detector $S_{\mathrm{SD}}$. The ratio 
between the observed and expected signals as a function of the polar 
angle on the ground in simulated showers is shown in the left panel of Fig.~\ref{tanksig}, with (thick 
points) and without (thin points) the geomagnetic field. While a 
significant quadrupolar modulation with a fixed phase along $D_\mathrm{B}$ and
amplitude $\simeq(1.1\pm 0.2)\%$ 
is observed when the field is on, no such modulation is observed when 
the field is off ($\simeq(0.1\pm 0.2)\%$), as expected. 
In the right panel, the same analysis is 
performed on the real data above 4~EeV, including again about $30\,000$ showers. A significant modulation of 
$\simeq(1.2\pm 0.2)\%$ is observed, agreeing both in amplitude and phase 
within the uncertainties with the simulations performed \textit{in the presence} 
of the geomagnetic field. This provides clear hints of the influence of the 
geomagnetic field in the Auger data. 

Note that this analysis is restricted to surface detectors that are more than 1000~m away from the shower core.
This cut is motivated by Fig.~\ref{mumap}, showing that the quadrupolar amplitude is larger at large 
distances from the shower core. We further require the surface detectors to have signals larger 
than 4~VEM\footnote{VEM - Vertical Equivalent Muon - is the average charge corresponding to the Cherenkov light produced by a vertical and central through-going muon in the surface detector. It is the unit used in the evaluation of the signal recorded by the detectors~\cite{bertou}.}. This cut is a compromise between keeping good statistics and keeping trigger effects small.
Above 4~VEM the measured amplitude does not depend systematically on the signal strength cut. However a cut in the surface detector signals 
induces a statistical trigger bias because showers with upward signal fluctuations will trigger more readily. This explains the small 
discrepancy between real and Monte-Carlo data in terms of the global normalisation in Fig.~\ref{tanksig} which differs from 1 by $\sim$3\%. Cutting at larger signals reduces this discrepancy.

Most importantly, depending on the incoming direction, the quadrupolar asymmetry is also expected to affect
the shower size $S(1000)$ and thus the energy estimator as qualitatively described in Eq.~\eqref{eqn:rho2}. Consequently,
these effects are expected to modulate the estimated cosmic ray event rate at a given energy as a 
function of the incoming direction, and in particular to generate a 
North/South asymmetry in the azimuthal distribution\footnote{The 
convention we use for the azimuthal angle $\varphi$ is to define it
relative to the East direction, counterclockwise.}. Such an asymmetry is
also expected in the case of a \emph{genuine} large scale modulation of the
flux of cosmic rays. However related analyses of the azimuthal distribution 
are out of the scope of this paper, and we restrict ourselves in the 
rest of this article to present a comprehensive study of the 
geomagnetic distortions of the energy estimator. This will allow us
to apply the corresponding corrections in a forthcoming publication
aimed at searching for large scale anisotropies.

\section{Geomagnetic distortions of the energy estimator}
\label{sec:azim}

\subsection{Geomagnetic distortions of the shower size $S(1000)$}
\label{sec:azim1}

The toy model presented in Section~\ref{sec:geom} allows us to 
understand the main features of the influence of the geomagnetic field 
on the muonic component of extensive air showers. To get an accurate 
estimation of the distortions induced by the field on the shower size 
$S(1000)$ as a function of both the zenith and the azimuthal angles, we 
present here the results obtained by means of end-to-end simulations of 
proton-initiated showers generated with the AIRES program~\cite{aires} 
and with the hadronic interaction model QGSJET~\cite{qgsjet}. We have 
checked that the results obtained with the CORSIKA program~\cite{corsika} 
are compatible. We consider a fixed energy $E=5$~EeV and seven fixed 
zenith angles between $\theta=0^\circ$ and $\theta=60^\circ$. The 
dependency of the effect in terms of the primary mass and of the number 
of muons in showers as well as its evolution with energy are sources of 
systematic uncertainties. The influence of such systematics will be 
quantified in Section~\ref{sec:syst_mod}. Within our convention for the azimuthal angle, the azimuthal direction 
of the magnetic North is $\varphi_\mathrm{B}^\mathrm{N}=90^\circ-D_\mathrm{B}=87.4^\circ$. 
The zenith direction of 
the field is $\theta_\mathrm{B}=90^\circ-|I_\mathrm{B}|=54.8^\circ$. 
\begin{figure}[!t]
  \centering					 
  \includegraphics[width=7cm]{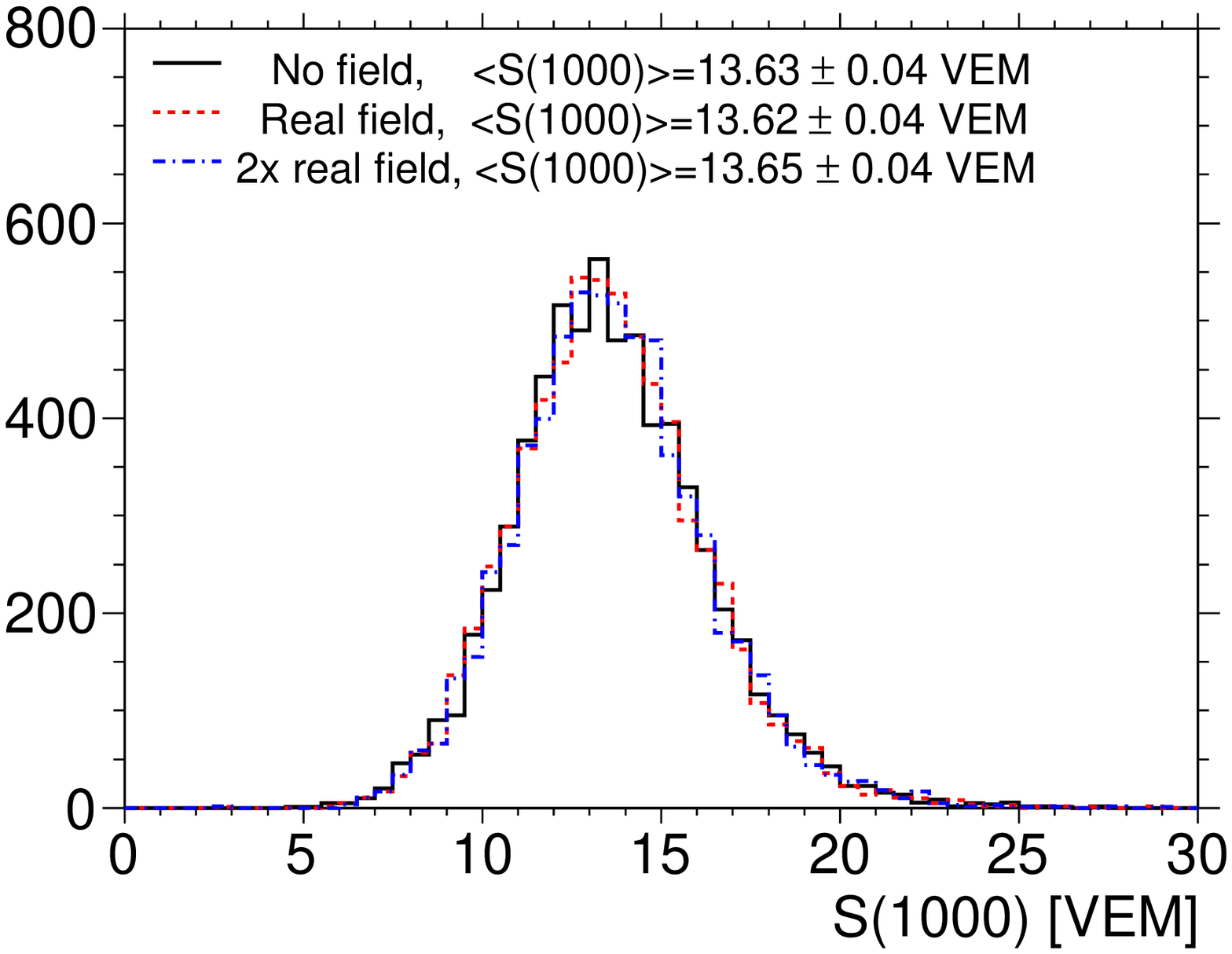}
  \includegraphics[width=7cm]{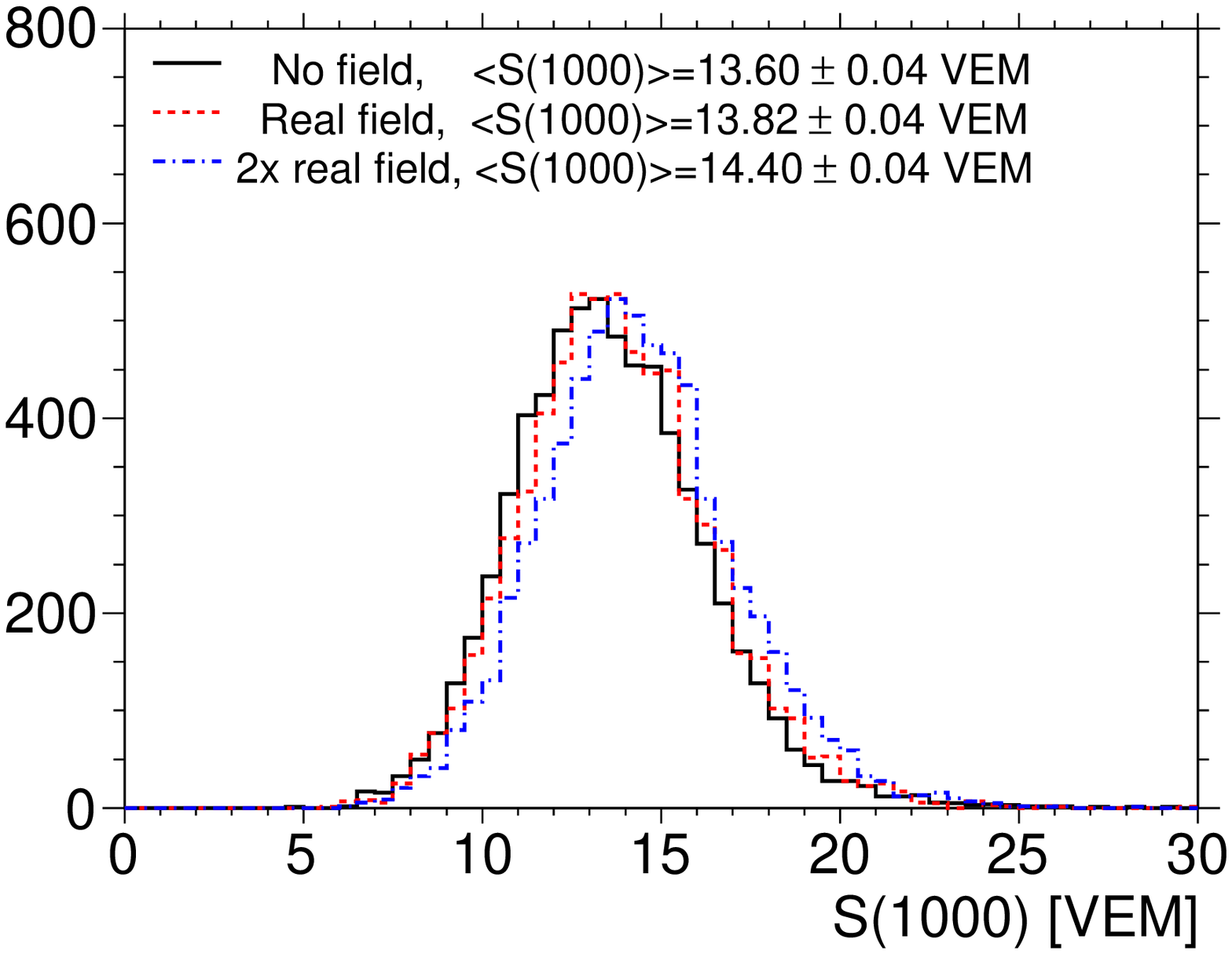}
  \caption{\small{Distributions of shower size $S(1000)$ obtained by simulating
showers at zenith angle $\theta=\theta_\mathrm{B}$ and azimuthal angle $\varphi_\mathrm{B}^\mathrm{N}$ 
(left) and $\varphi_\mathrm{B}^\mathrm{N}+90^\circ$ (right). Thick histogram: no magnetic
field. Dotted histogram: real magnetic field in Malarg\"{u}e. Dashed 
histogram: twice the real magnetic field in Malarg\"{u}e.}}
  \label{shiftS1000}
\end{figure}

To verify the predicted behaviour of the shower size shift in terms of
$B_{\rm T}^2$, we first show the results of the simulations of $1000$ showers
at a zenith angle $\theta=\theta_\mathrm{B}$ and for two distinct azimuthal angles 
$\varphi=\varphi_\mathrm{B}^\mathrm{N}$ and $\varphi=\varphi_\mathrm{B}^\mathrm{N}+90^\circ$. Each shower is 
then thrown 10 times at the surface detector array with random core 
positions and reconstructed using exactly the same reconstruction procedure 
as the one applied to real data. For this specific zenith angle $\theta_\mathrm{B}$, no shift 
is expected in the North direction $\varphi_\mathrm{B}^\mathrm{N}$ as the transverse component 
of the magnetic field is zero. This is indeed the case as illustrated 
in the left panel of Fig.~\ref{shiftS1000}, showing the distribution of 
reconstructed $S(1000)$ for three different configurations of the magnetic 
field: no field, real field in Malarg\"{u}e, and twice the real field in 
Malarg\"{u}e. It can be seen that on average all histograms are -- within the 
statistical uncertainties on the average -- centered on the same value. 
In the right panel of Fig.~\ref{shiftS1000} we repeat the same analysis 
with the showers generated in the direction $\varphi_\mathrm{B}^\mathrm{N}+90^\circ$. 
Since the transverse component of the field is now different from zero, a clear 
relative shift in terms of $\Delta S(1000)/S(1000)$ is observed between the three 
distributions: the shift is $\simeq 1.6\%$ between the configurations 
with and without the field, leading to a discrimination 
with a significance of $\simeq 5.5\,\sigma$, while the shift is 
$\simeq 6\%$ between the configurations with twice the real field and 
without the field leading to a discrimination with a significance of 
$\simeq 20\,\sigma$. It can be noticed that the strength of the shift is 
thus in overall agreement with the expected scaling $B_{\rm T}^2$.

\begin{figure}[!t]
  \centering					 
  \includegraphics[width=8cm]{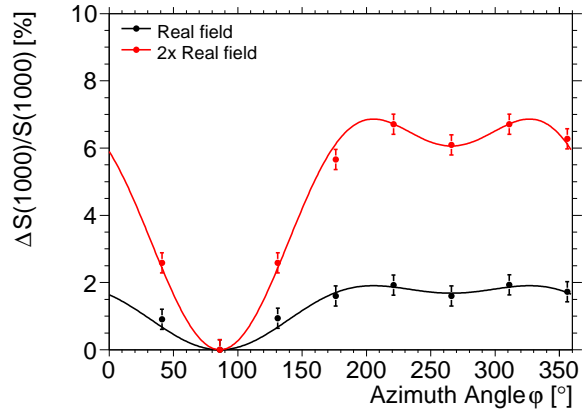}
  \caption{\small{$\Delta S(1000)/S(1000)$ (in \%) as a function of the 
azimuthal angle $\varphi$, at zenith angle $\theta=\theta_\mathrm{B}$ for two different field strengths. Points are obtained by Monte Carlo shower simulation, lines are the expected behavior (see Section~\ref{sec:geom}).}}
  \label{S1000vsPhi}
\end{figure}

For the zenith angle $\theta=\theta_\mathrm{B}$, in Fig.~\ref{S1000vsPhi} we show 
the shift of the mean $S(1000)$ obtained by simulating 1000 showers in the same way as previously for eight different values of the azimuth angle.
 Again, the results are displayed for configurations with the real
field (bottom) and with twice the real field (top). The expected behaviours
in terms of $\Delta S(1000)/S(1000)=G(\theta_\mathrm{B})\, \sin^2(\widehat{\textbf{u},\textbf{b}})$ are shown
by the continuous curves, where the normalisation factor $G$ is 
tuned by hand. Clearly, the shape of the curves agrees 
remarkably well with the Monte Carlo data within the uncertainties. Hence, this study supports the claim that 
the azimuthal dependence of the shift in $S(1000)$  induced by the 
magnetic field is proportional to $B_{\rm T}^2(\theta,\varphi)$, in agreement 
with the expectations provided by general considerations expressed in the 
previous section on the muonic component of the showers.

\begin{figure}[!t]
  \centering					 
  \includegraphics[width=8cm]{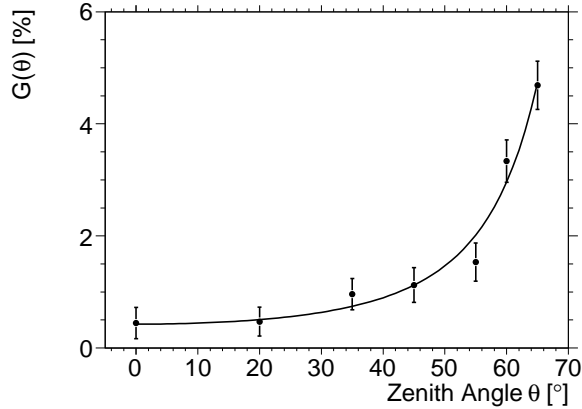}
  \caption{\small{$G(\theta)=\Delta S(1000)/S(1000)/\sin^2(\widehat{\textbf{u},\textbf{b}})$ as a function of the zenith angle $\theta$.}}
  \label{GvsTheta}
\end{figure}

The $B_{\rm T}^2$ term encompassing the overall azimuthal dependence at
each zenith angle, the remaining shift 
$G(\theta)=\Delta S(1000)/S(1000)/\sin^2(\widehat{\textbf{u},\textbf{b}})$ 
depends on the zenith angle through the altitude distribution of the muon 
production, the muon energy distribution, and the weight 
of the muonic contribution to the shower size $S(1000)$. Repeating 
the simulations at different zenith angles, we plot $G$ as a function of the zenith angle in Fig.~\ref{GvsTheta}. 
Due to the increased travel lengths of the muons and due to their larger relative 
contribution to $S(1000)$ at high zenith angles, the value of $G$ rises 
rapidly for angles above $\simeq 40^\circ$. The superimposed curve is an 
\emph{empirical} fit, allowing us to get the following parametrisation 
of the shower size distortions induced by the geomagnetic field,
\begin{equation}
\label{eqn:ds1000}
\frac{\Delta S(1000)}{S(1000)}(\theta,\varphi)=4.2\cdot 10^{-3}\, \cos^{-2.8}{\theta}\, \sin^2(\widehat{\textbf{u},\textbf{b}}).
\end{equation}

\subsection{From shower size to energy}
\label{sec:ener}

At the Pierre Auger Observatory, the shower size $S(1000)$ is
converted into energy $E$ using a two-step procedure~\cite{auger3}. 
First, the evolution of $S(1000)$ with zenith angle arising from 
the attenuation of the shower with increasing atmospheric thickness is quantified by 
applying the \emph{Constant Intensity Cut} (CIC) method that 
is based on the (at least approximate) isotropy of incoming cosmic
rays. The CIC relates relates $S(1000)$ in 
vertical and inclined showers through a line of equal intensity
in spectra at different zenith angles. This allows us to correct the value of $S(1000)$ for attenuation 
by computing its value had the shower arrived from a fixed zenith angle, here 38 degrees 
(corresponding to the median of the angular distribution of events for energies greater than 3~EeV).  This zenith angle independent estimator $S_{38}$ 
is defined as $S_{38}=S(1000)/CIC(\theta)$. The calibration of $S_{38}$ 
with energy $E$ is then achieved using a relation of the form 
$E=AS_{38}^B$, where $A=1.49\pm 0.06$(stat)$\pm 0.12$(syst) and 
$B=1.08\pm0.01$(stat)$\pm0.04$(syst) were estimated from the
correlation between $S_{38}$ and $E$ in a subset of high quality "hybrid events" measured simultaneously by the surface detector (SD) and the fluorescence detector (FD)~\cite{auger3}.  In such a sample, $S_{38}$ and E are independently measured, with $S_{38}$ from the SD and $E$ from the FD.

This two-step procedure has an important consequence on the 
implementation of the energy corrections for the geomagnetic
effects. The CIC curve is constructed assuming that the shower size estimator $S(1000)$ does not depend on the azimuthal angle. 
The induced azimuthal variation of $S(1000)$ due to the geomagnetic effect is thus 
averaged while the zenith angle dependence of the geomagnetic effects is absorbed when the CIC is implemented. To illustrate this in a simplified way, consider the case in which the magnetic field were directed along the zenith direction
 (\emph{i.e.} in the case of a virtual Observatory
located at the Southern magnetic pole) so that the transverse 
component of the magnetic field would not depend on the azimuthal 
direction of any incoming shower. Then the shift in $S(1000)$ 
would depend \emph{only} on the zenith angle in such a way that 
the Constant Intensity Cut method would by construction absorb the
shift induced by $G(\theta)$ into the empirical $CIC(\theta)$ curve,
while the empirical relationship $E=AS_{38}^B$ would calibrate
$S_{38}$ into energy with no need for any additional corrections.

This leads us to implement the energy 
corrections for geomagnetic effects, relating the energy $E_0$ reconstructed ignoring the geomagnetic effects to the \emph{corrected} energy $E$ by
\begin{equation}
\label{eqn:ecorr}
E=\frac{E_0}{(1+\Delta(\theta,\varphi))^B},
\end{equation}
with
\begin{equation}
\label{eqn:deltadef}
\Delta(\theta,\varphi) = G(\theta)\left[\sin^2(\widehat{\textbf{u},\textbf{b}})-\left<\sin^2(\widehat{\textbf{u},\textbf{b}})\right>_\varphi\right]
\end{equation}
where $\left<\cdot\right>_\varphi$ denotes the average over $\varphi$ and where $B$ is one of the parameters used in the $S_{38}$ to $E$ conversion described above. 
This expression implies that energies are \emph{under-estimated}
preferentially for showers coming from the northern directions of the
array, while they are \emph{over-estimated} for showers coming from
the southern directions, the size of the effect increasing with the zenith angle.

\section{Consequences for large scale anisotropy searches}
\label{sec:aniso}

\subsection{Impact on the estimated event rate}

\begin{figure}[!t]
  \centering					 
  \includegraphics[width=8cm]{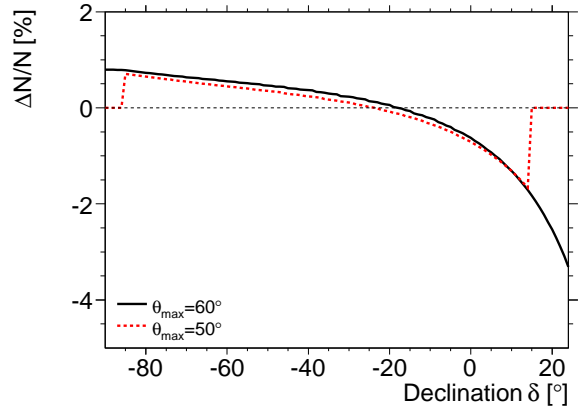}
  \caption{\small{Relative differences $\Delta N/N$ as a 
function of the declination, for 2 different values of $\theta_{\mathrm{max}}$.}}
  \label{deltaomega}
\end{figure}

To provide an illustration of the impact of the energy corrections 
for geomagnetic effects, we calculate here, as a function of declination $\delta$, the deviation of the event rate $N_0(\delta)$, 
measured if we were not to implement the corrections of the energy estimator by Eq.~\eqref{eqn:ecorr}, to the event rate $N(\delta)$ expected from an isotropic background distribution.

The ``canonical exposure''~\cite{sommers} holds
for a full-time operation of the surface detector array above the
energy at which the detection efficiency is saturated over the 
considered zenith range. In such a case, the directional detection
efficiency is simply proportional to $\cos{\theta}$,
\begin{equation}
\label{eqn:canon_omega}
\omega(\theta) \propto \cos (\theta) \, H(\theta - \theta_{\rm max})
\end{equation}
where $H$ is the Heaviside function and $\theta_{\mathrm{max}}$ is the maximal zenith
angle considered. The zenith angle is related to the declination $\delta$ 
and the right ascension $\alpha$ through
\begin{equation}
\label{eqn:costh}
\cos{\theta}=\sin{\ell_{\mathrm{site}}}\sin{\delta}+\cos{\ell_{\mathrm{site}}}\cos{\delta}\cos{\alpha}
\end{equation}
where $\ell_{\mathrm{site}}$ is the Earth's latitude of the Observatory.
The event rate at a given declination $\delta$ and above an energy threshold $E_{\rm th}$ is obtained by integrating in energy and right ascension $\alpha$,
\begin{equation}
\label{eqn:rate}
N(\delta) \propto \int_{E_{\rm th}}^\infty \mathrm{d}E \int_0^{2\pi}\mathrm{d}\alpha \, \omega(\theta) \, \frac{\mathrm{d}N(\theta,\varphi,E)}{\mathrm{d}E} 
\end{equation}
Note that at lower energies this integral acquires an additional energy and angle dependent detection efficiency term $\epsilon(E,\theta,\phi)$.
Hereafter we assume that the cosmic ray spectrum is a power law, \emph{i.e.} $dN/dE \propto E^{-\gamma}$. From Eq. (\ref{eqn:ecorr}) it follows that if the effect of the geomagnetic field were not accounted for, the measured energy spectrum would have a directional modulation given by
\begin{equation}
\label{eqn:spectrum2}
\frac{\mathrm{d}N}{\mathrm{d}E_0}\propto\left[1+\Delta(\theta,\varphi)\right]^{B(\gamma-1)}\, E_0^{-\gamma}.
\end{equation}
This leads to the following measured event rate above a given uncorrected energy $E_{\mathrm{th}}$,
\begin{equation}
\label{eqn:bias_omega2}
N_0(\delta)\propto \int_{E_{\mathrm{th}}}^\infty \mathrm{d}E_0 \int_0^{2\pi}\mathrm{d}\alpha\,H(\cos{\theta}-\cos{\theta_{\mathrm{max}}})\,\cos{\theta}\,\left[1+\Delta(\theta,\varphi)\right]^{B(\gamma-1)} E_0^{-\gamma},
\end{equation}
where $\varphi$ is related to $\alpha$ and $\delta$ through
\begin{equation}
\label{eqn:phi}
\tan{\varphi}=\frac{\sin{\delta}\cos{\ell_{\mathrm{site}}}-\cos{\delta}\cos{\alpha}\sin{\ell_{\mathrm{site}}}}{\cos{\delta}\sin{\alpha}}.
\end{equation}
The event rate $N_0(\delta)$ as a function of declination is then calculated using Eq. (\ref{eqn:spectrum2}) in Eq. (\ref{eqn:rate}). The relative difference $\Delta N/ N$ is shown in 
Fig.~\ref{deltaomega} as a function of the declination, with spectral index $\gamma=2.7$. 
The energy over-estimation (under-estimation) of events coming 
preferentially from the Southern (Northern) azimuthal directions, as 
described in Eq.~\eqref{eqn:ecorr}, leads to an effective excess (deficit) 
of the event rate for $\delta\lesssim -20^\circ$ ($\delta\gtrsim-20^\circ$),
with an amplitude of $\simeq 2\%$ when considering $\theta_{\mathrm{max}}=60^\circ$.
It is worth noting that this amplitude is reduced to within 1\% when
considering $\theta_{\mathrm{max}}=50^\circ$, as shown by the dotted line.

\subsection{Impact on dipolar modulation searches}

\begin{figure}[!t]
  \centering					 
  \includegraphics[width=7cm]{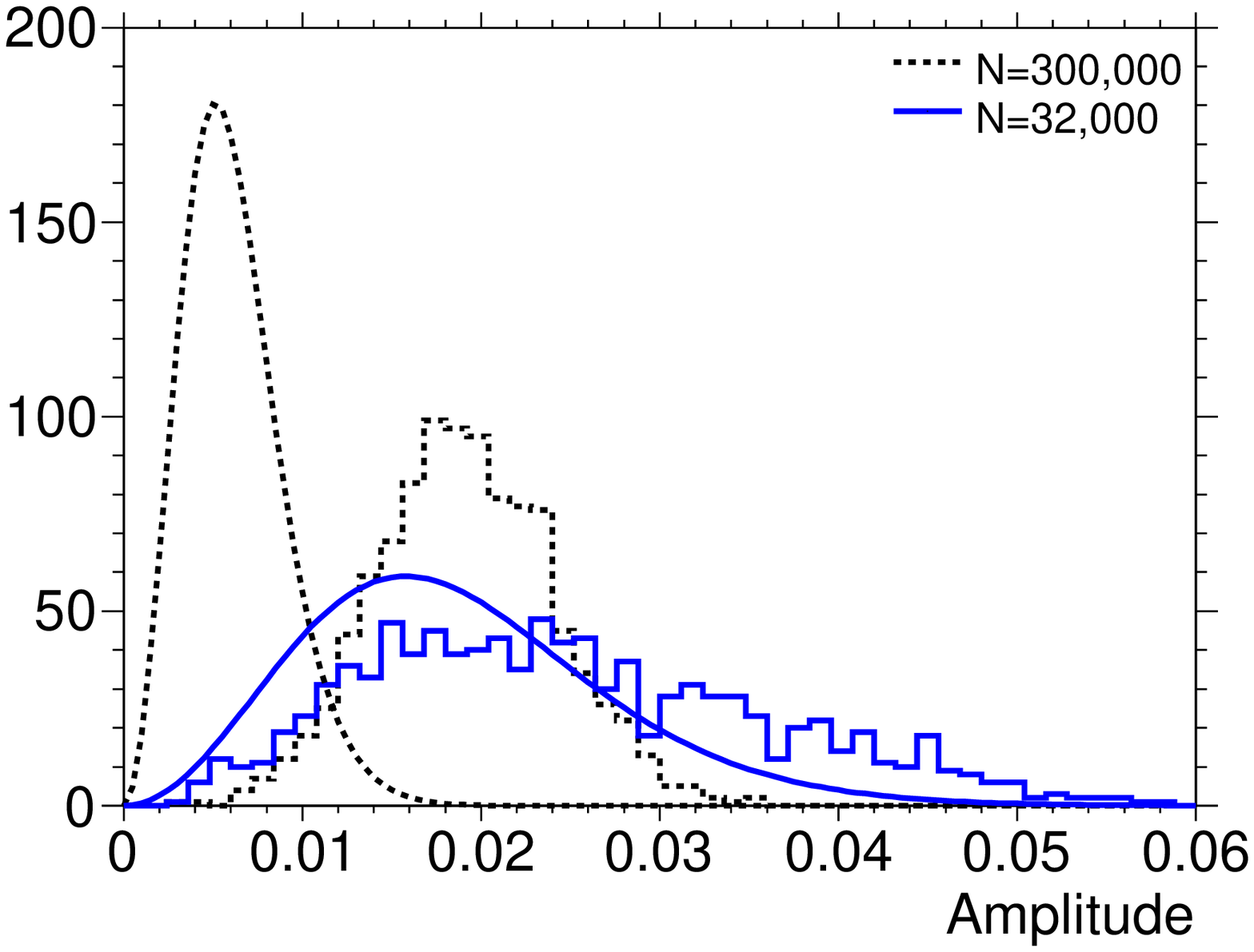}
  \includegraphics[width=7cm]{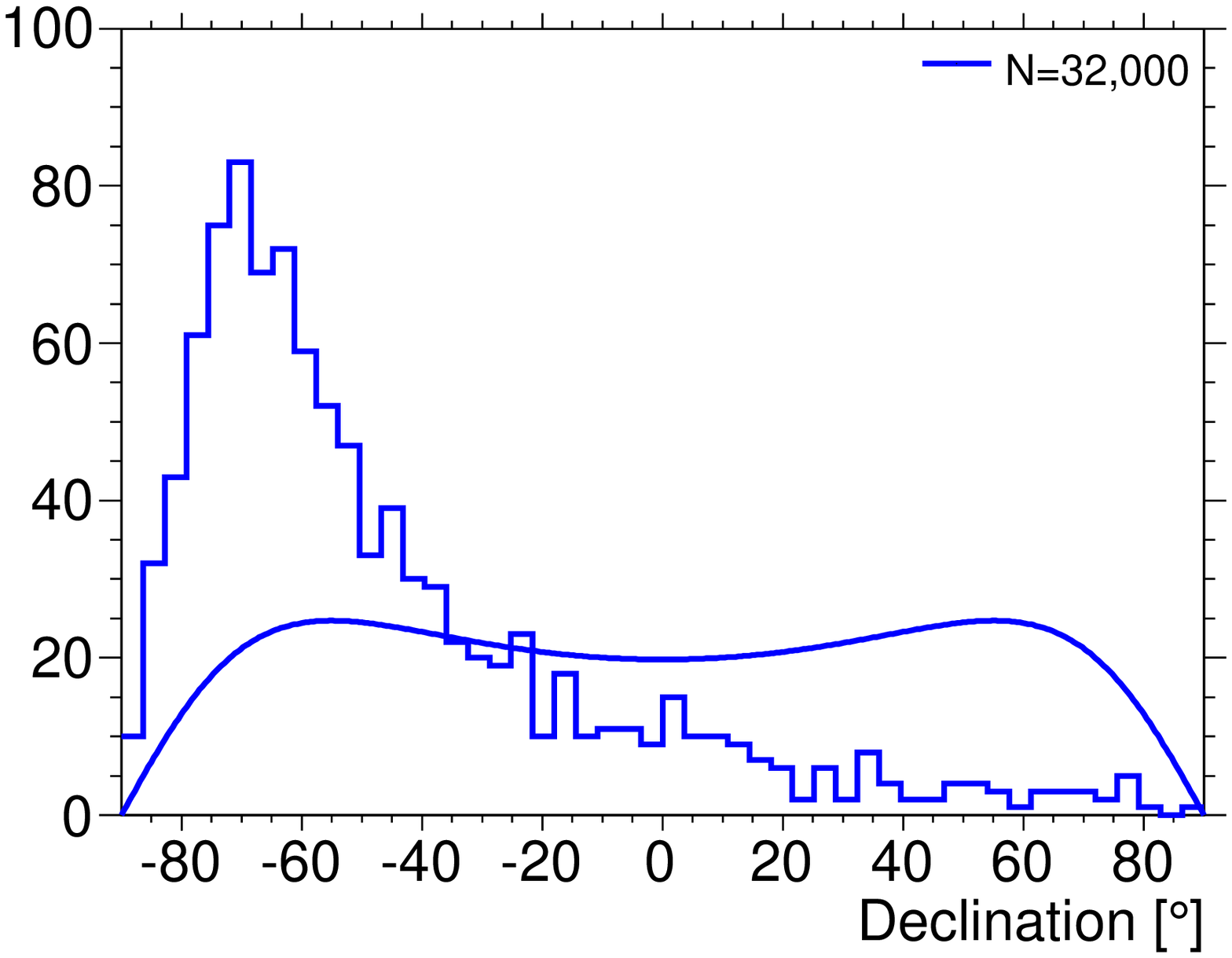}
  \caption{\small{Dipolar reconstruction of arrival directions of mock data 
sets with event rates distorted by the geomagnetic effects. Left: 
distributions of amplitudes. Right: distributions of declinations. The smooth lines give the expected distribution in the case of isotropy.}}
  \label{simus}
\end{figure}
The pattern displayed in Fig.~\ref{deltaomega} roughly imitates a dipole
with an amplitude at the percent level. To evaluate precisely 
the impact of this pattern on the assessment of a dipole moment in the reconstructed arrival directions and to probe
the statistics needed for the sensitivity to such a spurious 
pattern, we apply the multipolar reconstruction adapted 
to the case of a partial sky coverage~\cite{alm} to mock data sets by limiting the maximum 
bound of the expansion $L_{\mathrm{max}}$ to 1 (pure dipolar reconstruction). 
Since the distortions are axisymmetric around the axis defined by
the North and South celestial poles,
only the multipolar coefficient related to this particular axis is
expected to be affected (here: $a_{10}$). Consequently, this particular
coefficient has impacts on both the amplitude of the reconstructed
dipole and its direction with respect to the axis defined by
the North and South celestial poles (the technical details of relating the estimation of the multipolar coefficients 
to the spherical coordinates of a dipole are given in the Appendix).

To simulate the directional distortions induced by Eq.~\eqref{eqn:ecorr}, 
each mock data set is drawn from the event rate $N_0(\delta)$ corresponding to the uncorrected energies,
and is reconstructed using the canonical exposure in Eq.~\eqref{eqn:canon_omega}. 
The results of this procedure applied to $1000$ samples are shown in 
Fig.~\ref{simus}. In the left panel, the distribution of the reconstructed 
amplitudes $r$ using $N=300\,000$ events is shown by the dotted
histogram. It clearly deviates from the expected isotropic distribution 
displayed as the dotted curve which corresponds to (see Appendix)
\begin{equation}
\label{eqn:p_R}
p_R(r)=\frac{r}{\sigma\sqrt{\sigma_z^2-\sigma^2}}\,\mathrm{erfi}\bigg(\frac{\sqrt{\sigma_z^2-\sigma^2}}{\sigma\sigma_z}\frac{r}{\sqrt{2}}\bigg) \exp\bigg(-\frac{r^2}{2\sigma^2}\bigg),
\end{equation}
where $\mathrm{erfi}(z)=\mathrm{erf}(iz)/i$, and where the width parameters
$\sigma$ and $\sigma_z$ can be calculated from the exposure function~\cite{alm}.
With the particular exposure function used here, it turns out that
$\sigma\simeq1.02\sqrt{3/N}$ and $\sigma_z\simeq1.59\sqrt{3/N}$. This allows us to estimate the 
spurious dipolar amplitude\footnote{Due to the partial sky exposure 
considered here, the estimate of the dipolar amplitude is biased by the 
higher multipolar orders needed to fully describe $\Delta N/N$ 
shown in Fig.~\ref{simus}~\cite{alm}. The aim of this calculation is only 
to provide a quantitative illustration of the spurious measurement which 
would be performed due to the geomagnetic effects when reconstructing
a pure dipolar pattern.} to be of the order of the mean of the dotted 
histogram, about $\simeq 1.9\%$. Consequently, we can estimate that the 
spurious effect becomes predominant as soon as the mean noise amplitude 
$\left<r\right>$ deduced from Eq.~\eqref{eqn:p_R} is of the order 
of $1.9\%$,
\begin{equation}
\label{eqn:meanr}
\left<r\right>=\sqrt{\frac{2}{\pi}}\,\bigg(\sigma_z+\frac{\sigma^2\mathrm{arctanh}(\sqrt{1-\sigma^2/\sigma_z^2})}{\sqrt{\sigma_z^2-\sigma^2}}\bigg)\simeq 1.9\%.
\end{equation}
This translates into the condition $N\simeq 32\,000$ (solid histogram). Using such a number 
of events, the bias induced on the amplitude reconstruction is 
illustrated in the same graph by the longer tail of the full histogram 
with respect to the expected one, and is even more evident in the right 
panel of Fig.~\ref{simus}, showing the distribution of the reconstructed 
declination direction of the dipole which already deviates to a large 
extent from the expected distribution.

\section{Systematic uncertainties}
\label{sec:syst_mod}

\begin{figure}[!t]
  \centering					 
  \includegraphics[width=8cm]{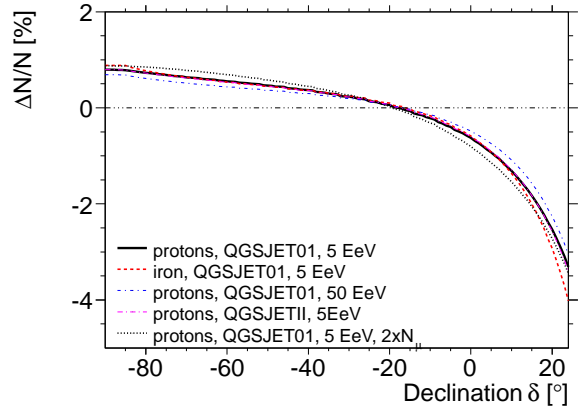}
  \caption{\small{Relative differences $\Delta N/N$ as a function 
of the declination, for different primary masses, different primary 
energies, different hadronic models and for increased number of muons 
in showers.}}
  \label{deltaomega_syst}
\end{figure}

The parametrisation of $G(\theta)$ in Eq.~\eqref{eqn:ds1000} was obtained
by means of simulations of proton showers at a fixed energy. The height
of the first interaction influences the production altitude of muons 
detected at $1000$~m from the shower core at the ground level. Moreover, as 
muons are produced at the end of the hadronic cascade, when the energy of 
the charged mesons is diminished so much that their decay length becomes 
smaller than their interaction length (which is inversely proportional to 
the air density), the energy distribution of muons is also affected by the 
height of the first interaction. Because the air density is lower in the 
upper atmosphere, this mechanism results in an increase of the energy of 
muons. The muonic contribution to $S(1000)$ depends also on both 
the primary mass and primary energy. For all these reasons, the 
parametrisation of $G(\theta)$ is expected to depend on both the primary 
mass and primary energy. 

To probe these influences, we repeat the same chain of end-to-end 
simulations using proton showers at energies of 50~EeV and iron showers 
at 5~EeV. Results in terms of the distortions of the observed event rate 
$N(\delta)$ are shown in Fig.~\ref{deltaomega_syst}. We also display 
in the same graph the results obtained using the hadronic interaction 
model QGSJETII~\cite{qgsjetII}. The differences with respect to the 
reference model are small, so that the consequences on large 
scale anisotropy searches presented in Section~\ref{sec:aniso} remain 
unchanged within the statistics available at the Pierre Auger 
Observatory.

In addition, there are discrepancies in the hadronic 
interaction model predictions regarding the number of muons in shower simulations and 
what is found in our data~\cite{ralph}. Higher 
number of muons influences the weight of the muonic contribution to 
$S(1000)$. The consequences of increasing the number of muons by a factor of 2 on the 
distortions of the observed event rate are also shown in Fig.~\ref{deltaomega_syst}. 
As the muonic contribution to $S(1000)$ is already large at high zenith 
angles in the reference model, this increase of the number of muons does 
not lead to large differences. 

\section{Conclusion}
\label{sec:conclusion}

In this work, we have identified and quantified a systematic uncertainty 
affecting the energy determination of cosmic rays detected by the surface 
detector array of the Pierre Auger Observatory. This systematic 
uncertainty, induced by the influence of the geomagnetic field on the 
shower development, has a strength which depends on both the zenith and 
the azimuthal angles. Consequently, we have shown that it induces 
distortions of the estimated cosmic ray event rate at a given energy at the percent level in both the 
azimuthal and the declination distributions, the latter of which mimics an almost 
dipolar pattern.

We have also shown that the induced distortions are already at the level
of the statistical uncertainties for a number of events $N\simeq 32\,000$
(we note that the full Auger surface detector array collects about
$6500$ events per year with energies above 3~EeV). 
Accounting for these effects is thus essential with regard to the correct 
interpretation of large scale anisotropy measurements taking explicitly 
profit from the declination distribution.

\section*{Acknowledgements}
The successful installation, commissioning, and operation of the Pierre Auger Observatory
would not have been possible without the strong commitment and effort
from the technical and administrative staff in Malarg\"ue.

We are very grateful to the following agencies and organizations for financial support: 
Comisi\'on Nacional de Energ\'ia At\'omica, 
Fundaci\'on Antorchas,
Gobierno De La Provincia de Mendoza, 
Municipalidad de Malarg\"ue,
NDM Holdings and Valle Las Le\~nas, in gratitude for their continuing
cooperation over land access, Argentina; 
the Australian Research Council;
Conselho Nacional de Desenvolvimento Cient\'ifico e Tecnol\'ogico (CNPq),
Financiadora de Estudos e Projetos (FINEP),
Funda\c{c}\~ao de Amparo \`a Pesquisa do Estado de Rio de Janeiro (FAPERJ),
Funda\c{c}\~ao de Amparo \`a Pesquisa do Estado de S\~ao Paulo (FAPESP),
Minist\'erio de Ci\^{e}ncia e Tecnologia (MCT), Brazil;
AVCR AV0Z10100502 and AV0Z10100522,
GAAV KJB100100904,
MSMT-CR LA08016, LC527, 1M06002, and MSM0021620859, Czech Republic;
Centre de Calcul IN2P3/CNRS, 
Centre National de la Recherche Scientifique (CNRS),
Conseil R\'egional Ile-de-France,
D\'epartement  Physique Nucl\'eaire et Corpusculaire (PNC-IN2P3/CNRS),
D\'epartement Sciences de l'Univers (SDU-INSU/CNRS), France;
Bundesministerium f\"ur Bildung und Forschung (BMBF),
Deutsche Forschungsgemeinschaft (DFG),
Finanzministerium Baden-W\"urttemberg,
Helmholtz-Gemeinschaft Deutscher Forschungszentren (HGF),
Ministerium f\"ur Wissenschaft und Forschung, Nordrhein-Westfalen,
Ministerium f\"ur Wissenschaft, Forschung und Kunst, Baden-W\"urttemberg, Germany; 
Istituto Nazionale di Fisica Nucleare (INFN),
Ministero dell'Istruzione, dell'Universit\`a e della Ricerca (MIUR), Italy;
Consejo Nacional de Ciencia y Tecnolog\'ia (CONACYT), Mexico;
Ministerie van Onderwijs, Cultuur en Wetenschap,
Nederlandse Organisatie voor Wetenschappelijk Onderzoek (NWO),
Stichting voor Fundamenteel Onderzoek der Materie (FOM), Netherlands;
Ministry of Science and Higher Education,
Grant Nos. N N202 200239 and N N202 207238, Poland;
Funda\c{c}\~ao para a Ci\^{e}ncia e a Tecnologia, Portugal;
Ministry for Higher Education, Science, and Technology,
Slovenian Research Agency, Slovenia;
Comunidad de Madrid, 
Consejer\'ia de Educaci\'on de la Comunidad de Castilla La Mancha, 
FEDER funds, 
Ministerio de Ciencia e Innovaci\'on and Consolider-Ingenio 2010 (CPAN),
Xunta de Galicia, Spain;
Science and Technology Facilities Council, United Kingdom;
Department of Energy, Contract Nos. DE-AC02-07CH11359, DE-FR02-04ER41300,
National Science Foundation, Grant No. 0450696,
The Grainger Foundation USA; 
ALFA-EC / HELEN,
European Union 6th Framework Program,
Grant No. MEIF-CT-2005-025057, 
European Union 7th Framework Program, Grant No. PIEF-GA-2008-220240,
and UNESCO.

\section*{Appendix}

The p.d.f. of the first harmonic amplitude for a data set of $N$ points 
drawn at random over a circle is known to be the Rayleigh distribution. 
In this appendix, we generalise this distribution to the case of $N$ points
being drawn at random on the sphere over the exposure $\omega(\delta)$ 
of the Pierre Auger Observatory. Assuming the underlying arrival direction 
distribution to be of the form 
$\Phi(\alpha,\delta)=\Phi_0(1+\textbf{D}\cdot\textbf{u})$, the components 
of the dipolar vector $\textbf{D}$ are related to the multipolar 
coefficients through
\begin{equation}
\label{eqn:coeffs}
D_x=\sqrt{3}\frac{a_{11}}{a_{00}}, \hspace{1cm} D_y=\sqrt{3}\frac{a_{1-1}}{a_{00}}, \hspace{1cm} D_z=\sqrt{3}\frac{a_{10}}{a_{00}}.
\end{equation}
Denoting by $x,y,z$ the estimates of $D_x,D_y,D_z$, the joint p.d.f. 
$p_{X,Y,Z}(x,y,z)$ can be factorised in the limit of large number of 
events in terms of three centered Gaussian distributions $N(0,\sigma)$,
\begin{equation}
\label{eqn:Pxyz}
p_{X,Y,Z}(x,y,z)=p_{X}(x)p_{Y}(y)p_{Z}(z)=N(0,\sigma_x)N(0,\sigma_y)N(0,\sigma_z),
\end{equation}
where the standard deviation parameters can be calculated
from the exposure function~\cite{alm}. With the particular exposure function
used here, it turns out that numerical integrations lead to
$\sigma\simeq1.02\sqrt{3/N}$ and $\sigma_z\simeq1.59\sqrt{3/N}$.
The joint p.d.f. $p_{R,\Delta,A}(r,\delta,\alpha)$ expressing the dipole components 
in spherical coordinates is obtained from Eq.~\eqref{eqn:Pxyz} by performing 
the Jacobian transformation
\begin{eqnarray}
\label{eqn:Prda}
p_{R,\Delta,A}(r,\delta,\alpha)&=&\left|\frac{\partial(x,y,z)}{\partial(r,\delta,\alpha)}\right| p_{X,Y,Z}(x(r,\delta,\alpha),y(r,\delta,\alpha),z(r,\delta,\alpha)) \nonumber \\
&=&\frac{r^2\cos\delta}{(2\pi)^{3/2}\sigma^2\sigma_z}\exp{\left[-\frac{r^2\cos^2\delta}{2\sigma^2}-\frac{r^2\sin^2\delta}{2\sigma_z^2}\right]}.
\end{eqnarray}
From this joint p.d.f., the p.d.f. of the dipole amplitude (declination) 
is finally obtained by marginalising over the other variables, yielding
\begin{eqnarray}
\label{eqn:Pr-d}
p_{R}(r)&=& \frac{r}{\sigma\sqrt{\sigma_z^2-\sigma^2}}\,\mathrm{erfi}\bigg(\frac{\sqrt{\sigma_z^2-\sigma^2}}{\sigma\sigma_z}\frac{r}{\sqrt{2}}\bigg) \exp\bigg(-\frac{r^2}{2\sigma^2}\bigg),\nonumber \\
p_{\Delta}(\delta)&=&\frac{\sigma\sigma_z^2}{2}\frac{\cos\delta}{(\sigma_z^2\cos^2\delta+\sigma^2\sin^2\delta)^{3/2}}.
\end{eqnarray}
Finally, one can derive from $p_R$ quantities of interest, such as the expected mean noise 
$\left<r\right>$, the RMS $\sigma_r$ and the probability of obtaining an
amplitude greater than $r$:
\begin{eqnarray}
\label{eqn:Pr-d2}
\left<r\right>&=&\sqrt{\frac{2}{\pi}}\,\bigg(\sigma_z+\frac{\sigma^2\mathrm{arctanh}(\sqrt{1-\sigma^2/\sigma_z^2})}{\sqrt{\sigma_z^2-\sigma^2}}\bigg), \\
\sigma_r&=&\sqrt{2\sigma^2+\sigma_z^2-\left\langle r \right\rangle^2}, \\
\mathrm{Prob}(>r)&=&\mathrm{erfc}\bigg(\frac{r}{\sqrt{2}\sigma_z}\bigg)+\frac{\sigma}{\sqrt{\sigma_z^2-\sigma^2}}\mathrm{ erfi}\bigg(\frac{\sqrt{\sigma_z^2-\sigma^2}}{\sqrt{2}\sigma\sigma_z}r\bigg)\mathrm{exp}\bigg(-\frac{r^2}{2\sigma^2}\bigg),
\end{eqnarray}
which are the equivalent to the well known Rayleigh formulas 
$\left<r\right>=\sqrt{\pi/N},\sigma_r=\sqrt{(4-\pi)/N}$ and 
$\mathrm{Prob}(>r)=\exp(-Nr^2/4)$ when dealing with $N$
points drawn at random over a circle~\cite{linsley}. 

\section*{Acknowledgments}

\end{document}